\pdfoutput=1

\documentclass[11pt]{article}

\usepackage[final]{acl}

\usepackage{times}
\usepackage{latexsym}

\usepackage[T1]{fontenc}

\usepackage[utf8]{inputenc}

\usepackage{microtype}

\usepackage{inconsolata}

\usepackage{graphicx}
\usepackage{amsmath,amssymb,amsfonts}
\usepackage{multirow,makecell}
\usepackage{bookmark}
\usepackage{booktabs}
\usepackage{url}
\usepackage{subfig}
\usepackage{xcolor,colortbl}
\newcommand{\greycell}{\cellcolor{gray!30}}

\newcommand{\nb}{\nolinebreak{}}

\newcommand{\bx}{\mathbf{x}}
\newcommand{\by}{\mathbf{y}}
\newcommand{\bs}{\mathbf{s}}

\usepackage{CJKutf8}
\newcommand{\jpchar}[1]{\begin{CJK}{UTF8}{ipxm}#1\end{CJK}}

\title{Integrating Pre-Trained Speech and Language Models \\ for End-to-End Speech Recognition}

\author{
  Yukiya Hono, {\bf Koh Mitsuda}, {\bf Tianyu Zhao}, \\ {\bf Kentaro Mitsui}, {\bf Toshiaki Wakatsuki}, {\bf Kei Sawada} \\ \\
  rinna Co., Ltd. \\
  \texttt{\{yuhono,kohmi,tianyuz,kemits,towaka,keisawada\}@rinna.co.jp}\\
}

\begin{document}
\maketitle
\begin{abstract}
  Advances in machine learning have made it possible to perform various text and speech processing tasks, such as automatic speech recognition (ASR), in an end-to-end (E2E) manner.
  E2E approaches utilizing pre-trained models are gaining attention for conserving training data and resources.
However, most of their applications in ASR involve only one of either a pre-trained speech or a language model.
  This paper proposes integrating a pre-trained speech representation model and a large language model (LLM) for E2E ASR.
  The proposed model enables the optimization of the entire ASR process, including acoustic feature extraction and acoustic and language modeling, by combining pre-trained models with a bridge network and also enables the application of remarkable developments in LLM utilization, such as parameter-efficient domain adaptation and inference optimization.
  Experimental results demonstrate that the proposed model achieves a performance comparable to that of modern E2E ASR models by utilizing powerful pre-training models with the proposed integrated approach.
\end{abstract}

\section{Introduction}

Mainstream of automatic speech recognition (ASR) has shifted from traditional pipeline methods to end-to-end (E2E) ones~\citep{graves2013speech,chorowski2015attention}.
Some of the key techniques in E2E ASR include a connectionist temporal classification (CTC)~\citep{graves2013speech}, and an attention-based encoder-decoder model~\citep{chorowski2015attention}.
Alternatively, speech representation models trained in a self-supervised manner using large amounts of unlabeled speech data have also attracted considerable attention~\citep{baevski2020wav2vec2,hsu2021hubert}.
Fine-tuning these models achieves a better performance in downstream tasks, including ASR, particularly in low-data settings.
However, these methods typically employ external language model-fused decoding, which complicates the decoding process.
Although a more recent study~\citep{radford2023robust} showed remarkable success with a single encoder-decoder model using 680k hours of weakly supervised labeled data without such pre-training models and decoding fusion, it is challenging to collect data on this scale.

In natural language processing (NLP) research, large language models (LLMs) pre-trained on massive amounts of text data have significantly impacted NLP benchmarks, such as question answering, knowledge retrieval, and natural language understanding~\citep{radford2018improving,black2022gpt,chowdhery2022palm,touvron2023llama}.
In recent years, the success of LLMs has sparked research interest in integrating LLMs with other modalities such as images, audio, and video~\citep{alayrac2022flamingo,li2023blip,zhu2023minigpt,chang2023speechprompt,zhang2023speechgpt,shen2023hugginggpt,wang2023viola,rubenstein2023audiopalm,fathullah2023prompting,wu2023decoder,pan2023cosmic,yu2023connecting,chen2023x}.
Pioneering studies~\citep{zhang2023speechgpt,shen2023hugginggpt} utilized LLM as a control hub with several well-trained speech expert models.
Several studies have investigated feeding the audio information to the LLM as a prompt so that the decoder-only Transformer~\citep{vaswani2017attention} can adapt to speech processing tasks~\citep{chang2023speechprompt,zhang2023speechgpt,wang2023viola,rubenstein2023audiopalm,fathullah2023prompting,wu2023decoder,pan2023cosmic,yu2023connecting}.
These methods demonstrate superior performance in various speech-text processing tasks such as ASR and automatic speech translation, benefiting from the powerful linguistic knowledge of LLMs.
In addition, some work on ASR and speech synthesis have also shown the potential of a pure decoder-only Transformer that has yet to be pre-trained~\citep{tsunoo2023decoder,wang2023neural}.

Inspired by recent successes in speech representation learning and LLMs, this paper proposes a novel E2E ASR model by integrating two large-scale pre-trained models, HuBERT~\citep{hsu2021hubert} and GPT~\citep{radford2018improving}.
Our proposed method benefits from the powerful language modeling and decoding capabilities of autoregressive LLMs, eliminating the need for complex decoding processes, such as external language model-fused decoding.
Even with simple greedy decoding, our approach achieves a performance comparable to that of recent well-trained and carefully designed ASR models.
Furthermore, our model can take advantage of recent rapid developments in LLM research fields, such as inference optimization tools and rich domain adaptation knowledge, which suggest promising avenues for future progress.
The primary contributions of this paper are as follows:
\begin{itemize}
  \item We propose a fully E2E ASR model capable of directly generating text token sequences from speech waveforms, integrating pre-trained HuBERT and GPT models.
  \item Through ablation tests, we identify the optimal way to integrate these two pre-trained models.
  \item We demonstrate the fine-tuning of the proposed model for domain adaptation, which will be helpful for insight into the deployment of the proposed model in practical use cases.
\end{itemize}
We call the proposed model \textit{Nue-ASR}\footnote{The name comes from the Japanese word ``Nue,'' one of the Japanese legendary creatures ``Yōkai.''} and release an inference code with the model checkpoint for Japanese E2E ASR\footnote{\url{https://huggingface.co/rinna/nue-asr}}.

\section{Related Work}

There has been a recent surge in the use of LLMs for speech processing tasks, including ASR.
This emerging trend reflects a growing awareness of the potential of LLMs to apply their ability to perform a variety of tasks.

The primary approach to directly exploiting LLMs for speech processing tasks is to feed the continuous representation obtained from the speech processing model~\citep{fathullah2023prompting,wu2023decoder,pan2023cosmic,yu2023connecting} or discretized features from its representation~\citep{chang2023speechprompt,zhang2023speechgpt,wang2023viola,rubenstein2023audiopalm,chen2023lauragpt} to the LLM, empowering the decoder-only Transformer to recognize and generate multimodal content, including audio.
In practice, continuous speech representations or discrete audio tokens are injected into a linguistic embedding space.

In the continuous representation-based approach, Transformer-based speech encoders are often introduced to obtain speech representations from acoustic features such as Mel filterbank outputs.
The sequence length of these features is longer than that of text tokens since these kinds of features are often 50 and 100 frames per second.
Therefore, downsampling using a convolution layer~\citep{fathullah2023prompting} or CTC compression~\citep{gaido2021ctc,wu2023decoder,pan2023cosmic} is often employed.
Other approaches employ a Querying Transformer (Q-Former)~\citep{li2023blip} to convert variable-length speech representations into fixed-length output query representations, or an enhancement of the Q-Former to extract segment-level query representations~\citep{yu2023connecting}.
These approaches involve pre-training the speech encoder module or CTC compression network with paired speech and text data using CTC loss, or leveraging the encoder from a well-trained encoder-decoder ASR model.
This implies a substantial requirement for training the speech processing model with ASR tasks.

In the discrete representation-based approach, the outputs of speech representation models such as HuBERT~\citep{hsu2021hubert} and w2v-BERT~\citep{chung2021w2v} are discretized via $k$-means, or neural audio codec models such as EnCodec~\citep{defossez2022high} are utilized to extract discrete token from speech waveforms.
To address the gap between the sequence lengths of speech and text tokens, a technique has also been proposed to remove adjacent duplicate indices.
Discrete speech tokens can be treated similarly to text tokens by extending the LLM vocabulary.
However, detailed information in speech may not be fully exploited, and optimization is inherently a two-step process.

This paper integrates a pre-trained speech representation model with the LLM to utilize the speech representations extracted from speech waveforms as continuous features directly.
Our method, categorized as a continuous representation-based approach, diverges from prior studies by not relying on signal processing-based acoustic features, thus completely enjoying the advantages of a speech representation model with self-supervised learning.
The integrated model with a convolution-based bridge network can be trained with only a single causal language modeling objective.
Consequently, it achieves fully E2E ASR by directly processing speech waveforms, exploiting the combined capabilities of speech models and LLMs.
This paper focuses on the ASR task, which is appropriate as a first consideration for integrating pre-trained speech and language models for E2E speech processing tasks.
It can assess whether speech information is accurately passed to the LLM without missing any content.

\section{Methodologies}

\begin{figure}[t]
  \centering
  \includegraphics[width=0.95\linewidth]{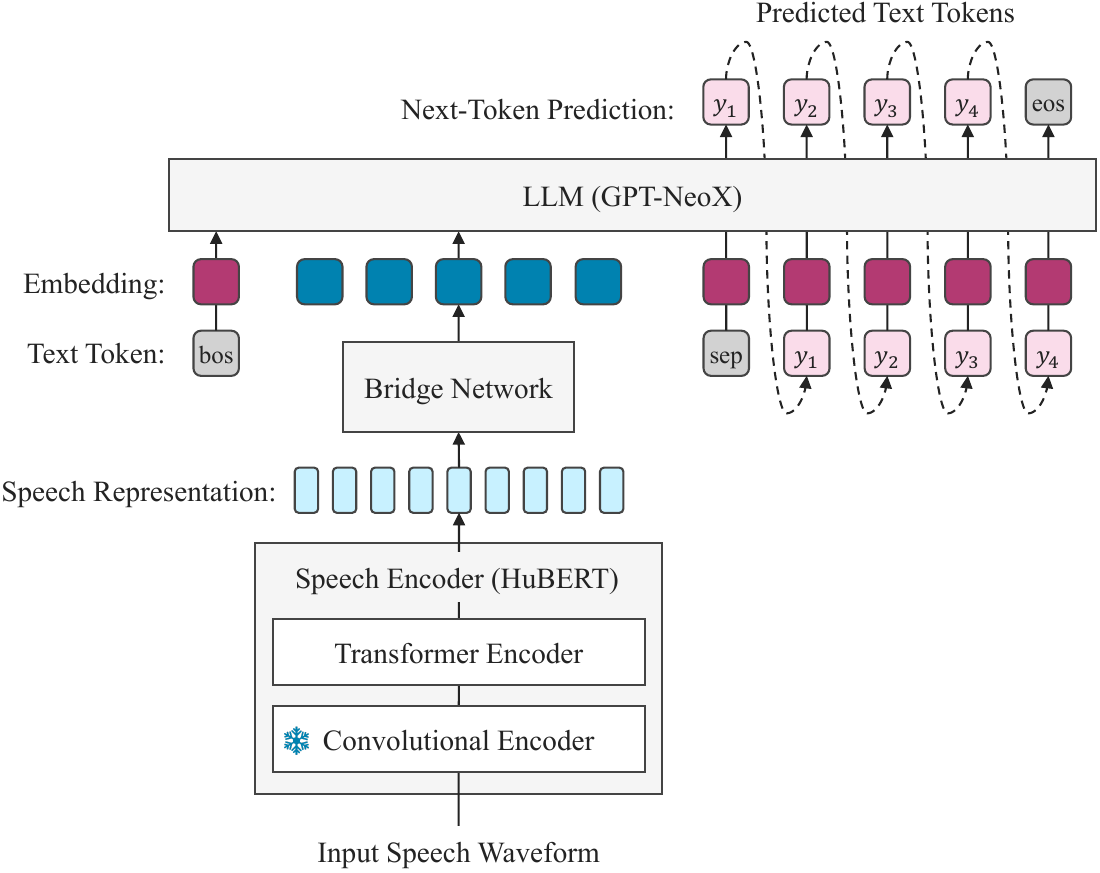}
  \caption{
    Overview of the proposed model.
    All modules of the speech encoder, bridge network, and LLM, except the convolutional waveform encoder, are simultaneously optimized in an E2E manner.
  }
  \label{fig:overview}
  \vspace{-1mm}
\end{figure}

An overview of the proposed model is presented in Fig~\ref{fig:overview}.
This model consists of three main components: speech encoder, bridge network, and LLM.

\subsection{Leveraging LLMs for speech recognition}

The proposed model uses a decoder-only Transformer-based LLM to perform speech recognition as next-token prediction.
Any LLM can be used.
We adopted the pre-trained GPT-NeoX~\citep{andoniangpt2021gpt}, a GPT variant with a modified architecture for the Transformer layer and an alternative positional encoding mechanism called rotary embedding~\citep{su2021roformer} to replace the original learnable position embeddings.
In practice, a speech waveform $\bx$ is fed into an speech encoder to obtain speech representations, followed by a bridge network to convert the speech representations into the embedding space of text tokens to feed LLM as speech prompts.
Given speech prompt $\bs$, the generation of the corresponding text sequence $\by$ is formulated as follows:
\begin{align}
  p(\by|\bs; \theta_\mathrm{LM}) = \prod_{i=1}^I p(y_i|\by_{<i}, \bs; \theta_\mathrm{LM}),
\end{align}
where $\by_{<i}$ is the generated text sequence before time step \nb$i$, $I$ is the number of tokens contained in the text sequence, and $\theta_\mathrm{LM}$ denotes the parameters of the LLM.

\subsection{Prompting speech information}
\label{sec:prompt}

We adopted a pre-trained HuBERT model as the speech encoder, focusing on the potential of self-supervised speech representation learning with large amounts of speech data.
HuBERT contains a convolutional waveform encoder and a BERT-like Transformer-based encoder~\citep{devlin2019bert}, that is trained in a self-supervised manner using a masked prediction objective~\citep{hsu2021hubert}.

\begin{figure}[t]
  \centering
  \includegraphics[width=0.95\linewidth]{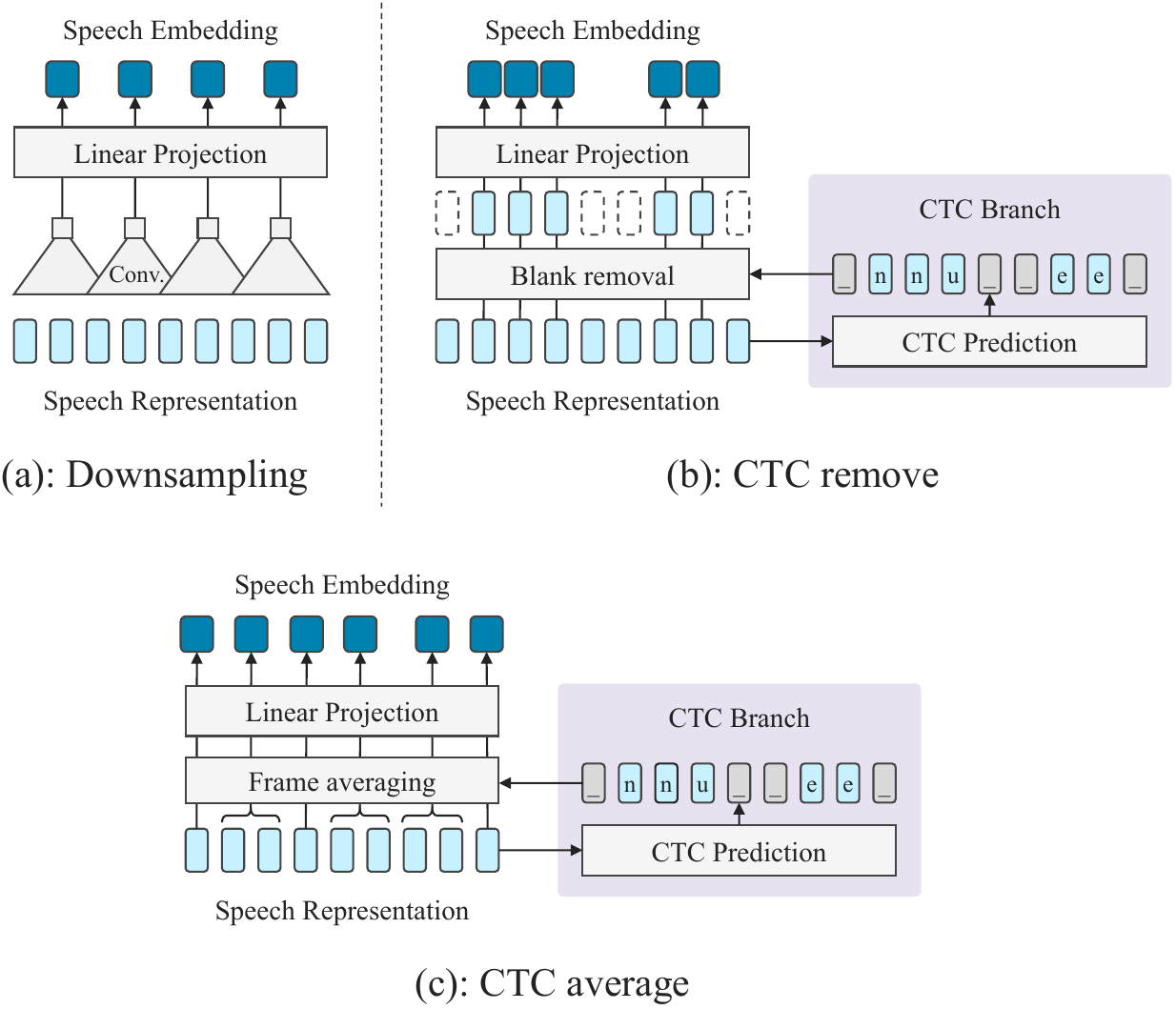}
  \caption{
    The details of the bridge network.
    In (b) CTC remove and (c) CTC average, a dedicated softmax layer is placed on top of the HuBERT encoder as a CTC branch. An additional CTC loss is also introduced.
  }
  \label{fig:bridge}
  \vspace{-1mm}
\end{figure}

The output of the speech encoder is fed into the bridge network, which converts the speech representations into the embedding space of the LLM.
Because the output of HuBERT is a 20 ms shifted feature sequence, the sequence length is longer than that of text tokens, making direct handling by the LLM inefficient.
Therefore, sequence compression is performed when a bridge network maps HuBERT features onto the LLM embedding space.
Several studies~\citep{gaido2021ctc,wu2023decoder,pan2023cosmic,tsunoo2023decoder} introduces CTC compression based on the tokens predicted by CTC~\citep{graves2013speech}.
Motivated by the success of previous studies, the following three methods are considered for sequence compression in bridge networks.
\begin{description}
  \item[Downsampling:]
  The output sequence from HuBERT is downsampled to a quarter of its original length using two convolutional layers with a kernel size of 4 and a stride of 2.
  \item[CTC remove:]
  The output frames of HuBERT that CTC predicts as blank are removed.
  CTC prediction is performed by adding a dedicated softmax layer on top of the HuBERT encoder as a CTC branch.
  \item[CTC average:]
  The adjacent output frames of HuBERT that CTC predicts as the same symbol are averaged.
  CTC prediction is performed in the same manner as CTC remove.
\end{description}
The details of the bridge network using these three methods are shown in Fig~\ref{fig:bridge}.
CTC remove and CTC average are denoted as CTC compression in the following sections.

\subsection{Training strategy}

The proposed model can be trained using a causal language modeling loss function as follows:
\begin{align}
  \mathcal{L}_\mathrm{LM} = -\sum_{i=1}^I \log p(y_i|\by_{<i}, \bs). \label{eq:llm_objective}
\end{align}
Furthermore, when CTC compression is applied within the bridge network, the model should be trained using CTC loss~$\mathcal{L}_\mathrm{CTC}$~\citep{graves2006connectionist} along with the causal language modeling loss, as follows:
\begin{align}
  \mathcal{L}_\mathrm{LM+CTC} = \mathcal{L}_\mathrm{LM} + \lambda_\mathrm{CTC} \mathcal{L}_\mathrm{CTC},
\end{align}
where $\lambda_\mathrm{CTC}$ is a hyperparameter to controlling the weight of CTC loss.

Intuitively, to maximize the performance of our proposed model, it would be desirable to train all parameters of the speech encoder, bridge network, and LLM simultaneously.
This holistic approach ensures complete integration and optimization across all components in a fully E2E manner.
However, because pre-trained models, especially LLM, have many parameters, full-parameter fine-tuning of these models is computationally expensive.
It is also worth considering the effectiveness of the built-in capabilities of these pre-trained models for ASR tasks.
Therefore, this paper will also consider cases where each model is frozen through experiments in Section~\ref{sec:exp_abl}.

As a popular technique for adapting large-scale LLMs to new datasets or tasks at a minimal cost, parameter efficient fine-tuning (PEFT)~\citep{houlsby2019parameter} has recently gained attention.
Most of the previous studies~\citep{zhang2023speechgpt,fathullah2023prompting,wu2023decoder,pan2023cosmic} to extend LLMs to speech processing tasks have use low-rank adaptation (LoRA)~\citep{hu2021lora} to perform PEFT.
LoRA allows for efficient adaptation by modifying a smaller subset of model's parameters, enabling task-specific adjustments without the extensive computational overhead typically associated with training large models.
Therefore, this study will also examine the optimization of an efficient LLM part using LoRA.

Through these methodologies, our study aims to create an efficient and effective bridge between speech and text modalities, harnessing the strengths of pre-trained speech and language models.

\section{Experiments}
\label{sec:exp}

\subsection{Setup}
\label{sec:exp_setup}

To evaluate the performance of the proposed model, we conducted Japanese ASR experiments using publicly available speech corpora.
We used the ReazonSpeech corpus (v1.0)~\citep{yin2023reazonspeech}, a 19,000-hour speech corpus collected from Japanese TV programs with a 16~kHz sampling.
We used 1,000 utterances as development data, 1,000 different utterances as test data, and the remainder as training data.
Before ASR training, all symbols except hyphens, apostrophes, punctuation, and question marks were removed as part of the text normalization process.
Further details of the dataset are provided in the Appendix~\ref{sec:appendix_data}.

To initialize the speech encoder, we used a pre-trained japanese-hubert-base\footnote{\url{https://huggingface.co/rinna/japanese-hubert-base}} model~\citep{sawada2024release}, which was trained on the ReazonSpeech corpus, consisting of 7-layer convolutions and a 12-layer Transformer with 768 hidden units and 12 attention heads.
To initialize the LLM, we used a pre-trained japanese-gpt-neox-3.6b\footnote{\url{https://huggingface.co/rinna/japanese-gpt-neox-3.6b}} model~\citep{sawada2024release}, which was trained on approximately 312.5 billion tokens, consisting of a 36-layer Transformer with 2,816 hidden units and 22 attention heads.
We used a SentencePiece~\citep{kudo2018sentencepiece}-based tokenizer.
The tokenizer had a vocabulary size of 32,000.
The proposed models were trained on four NVIDIA A100 80GB GPUs for a single run of five epochs with four gradient accumulations.
The total batch size was set to 64 utterances per GPU.
The optimizer was AdamW~\citep{loshchilov2019decoupled} with parameters $\beta_1=0.9$, $\beta_2=0.999$, and a weight decay factor of~$0.05$.
The learning rate was linearly warmed up over a quarter of the epoch up to a peak value of $0.0001$ followed by a cosine-decaying schedule.
The wall-time required to train the proposed model was approximately 56 hours.
For CTC compression, $\lambda_\mathrm{CTC}$ was set to 0.5.
To efficiently train the proposed model with CTC compression, HuBERT was fine-tuned in advance using only the CTC loss.
This CTC fine-tuned model is referred to as HuBERT-CTC hereinafter.
In the preliminary experiments, we found that the convolutional waveform encoder in HuBERT should be frozen for stable training of the proposed model.

We optimized the inference process of the GPT in the proposed model using DeepSpeed-Inference~\citep{aminabadi2022deepspeed}.
Greedy search was used for inference the proposed model.
We investigated suitable decoding strategies for the proposed model in Appendix~\ref{sec:appendix_decode}.

\subsection{Ablation study}
\label{sec:exp_abl}

The performance of the proposed model was compared under various conditions.
In addition to the ReazonSpeech test set, we used two other Japanese speech corpora for evaluation: 1) \nb JSUT basic5000~\citep{sonobe2017jsut}, which contains 5,000 utterances by a single female speaker with high-quality recordings for speech synthesis, and 2) Common Voice 8.0 (CV8.0) Japanese test set~\citep{ardila2020common}, which consists of 4,483 utterances by multiple speakers.
The detailed information of additional datasets is provided in Appendix~\ref{sec:appendix_data}.

As introduced in Section~\ref{sec:prompt}, we compared three methods of bridging the HuBERT output space and the GPT input embedding space: Downsampling with convolution, CTC remove, and CTC average.
In each case, we compared whether to fine-tune the HuBERT encoder and GPT.
Following previous studies~\citep{fathullah2023prompting}, we investigated using LoRA to efficiently fine-tune GPT parameters with a rank parameter $R=32$ instead of full-parameter fine-tuning (Full FT).
We also included a method that uses only HuBERT-CTC in this comparison.
We measured ASR performance by calculating character error rates (CERs) for text by removing all symbols, including punctuation marks, and converting numbers to words using num2words\footnote{\url{https://github.com/savoirfairelinux/num2words}}.

\begin{table}[t]
  \centering
  \resizebox{0.93\linewidth}{!}{
    \begin{tabular}[t]{cccccc}
      \toprule
      \textbf{\multirow{2}{*}{IDs}} & \multicolumn{2}{c}{\textbf{Model description}} & \multicolumn{3}{c}{\textbf{CER [\%] ($\downarrow$)}} \\
      \cmidrule(lr){2-3} \cmidrule(lr){4-6}
      & \textbf{HuBERT} & \textbf{GPT} & \textbf{Reazon} & \textbf{JSUT} & \textbf{CV8.0} \\
      \midrule\midrule
      \multicolumn{6}{l}{\hspace{-0.6em} \textbf{HuBERT + GPT (w/ Downsampling)}} \\
      A1 & Frozen  & Full FT & 9.7  & 10.4 & 10.2 \\
      A2 & Full FT & Frozen  & 10.8 & 19.1 & 13.4 \\
      A3 & Full FT & Full FT & 8.4  & \textbf{8.6} & \textbf{9.2} \\
      A4 & Full FT & PEFT    & 10.4 & 14.8 & 12.1 \\
      \midrule
      \multicolumn{6}{l}{\hspace{-0.6em} \textbf{HuBERT-CTC + GPT (w/ CTC remove)}} \\
      B1 & Frozen  & Full FT & 9.1  & 12.1 & 11.9 \\
      B2 & Full FT & Frozen  & 13.6 & 18.5 & 18.4 \\
      B3 & Full FT & Full FT & \textbf{7.8} & 12.1 & 10.4 \\
      B4 & Full FT & PEFT    & 10.2 & 21.6 & 14.6 \\
      \midrule
      \multicolumn{6}{l}{\hspace{-0.6em} \textbf{HuBERT-CTC + GPT (w/ CTC average)}} \\
      C1 & Frozen  & Full FT & 8.6  & 11.9 & 10.6 \\
      C2 & Full FT & Frozen  & 10.5 & 17.0 & 15.1 \\
      C3 & Full FT & Full FT & \textbf{7.8} & 11.5 & 9.7 \\
      C4 & Full FT & PEFT    & 10.1 & 15.3 & 13.0 \\
      \midrule
      \multicolumn{6}{l}{\hspace{-0.6em} \textbf{HuBERT-CTC}} \\
      D  & Frozen & - & 16.3 & 23.8 & 21.5 \\
      \bottomrule
    \end{tabular}
  }
  \caption{ASR results of ablation test with ReazonSpeech test set, JSUT basic5000, and Common Voice 8.0 (CV8.0) test set.}
  \label{table:exp_abl}
  \vspace{-1mm}
\end{table}

The results are presented in Table~\ref{table:exp_abl}.
All models integrating HuBERT and GPT (A1--4, B1--4, C1--4) significantly outperformed HuBERT-CTC (D).
The results show that integrating HuBERT with GPT, which has vast linguistic knowledge, improves recognition accuracy.
In addition, fine-tuning the full parameters of both HuBERT and GPT (A3, B3, C3) achieved lower CERs, regardless of the bridge network type.
We also observed that GPT fine-tuning was more effective than HuBERT fine-tuning.
When GPT was frozen (A2, B2, C2), the bridge network was forced to map the speech representation to the LLM embedding space suitable for text tokens to perform the ASR task via GPT in a zero-shot manner.
By contrast, GPT fine-tuning enhanced the ability to handle speech information directly and appropriately.
When HuBERT was frozen (A1, B1, C1), the recognition accuracy decreased only slightly because GPT could be optimized to handle speech information.
This also implies that pre-trained HuBERT and HuBERT-CTC already have the potential to extract useful features for ASR.

In evaluating the system that performed PEFT with LoRA on GPT (A4, B4, C4), we observed that its performance exceeded that of the corresponding system with a frozen GPT (A2, B2, C2).
However, it did not reach the level of systems that had undergone full-parameter fine-tuning (A3, B3, C3).
These results suggest that GPT optimization significantly contributes to achieving high recognition accuracy in the proposed model.

Here, we compare sequence compression techniques in bridge networks among models using the same training strategies.
When comparing the CTC-based compression methods, CTC average groups (C1--3) demonstrated better performance in all training strategies than CTC remove groups (B1--3), particulally for the JSUT basic5000 and CV8.0 test sets.
This result differs from previous ASR studies that employed a decoder-only Transformer model~\citep{tsunoo2023decoder}, which could be attributed to whether or not pre-training models were used or the fact that a different target language may have affected the results.
In contrast, Downsampling (A1--3) showed different trends for different test sets.
CTC average (C1--3) showed improved performance on the ReazonSpeech test set, yet Downsampling (A1-3) consistently outperformed CTC on JSUT basic5000 and CV8.0, marking a departure from prior findings~\citep{tsunoo2023decoder,wu2023decoder}.
The results suggest that CTC-based compression is sensitive to the domain match between the training and evaluation data.
Although CTC compression can effectively compress the sequence of speech representations to be close to the text length, CTC compression's inaccuracies, often due to prediction errors, can negatively impact GPT-driven text generation.
In addition, allowing the speech encoder to perform CTC prediction implies a shift to a two-stage model, potentially compromising end-to-end optimization compared to downsampling.
A detailed analysis can be found in Appendix~\ref{sec:appendix_abl}.
We also investigated the utilizing different speech encoder in Appendix~\ref{sec:appendix_encoder}.

\subsection{Comparison with other ASR models}
\label{sec:exp_bench}

\begin{table*}[t]
  \centering
  \resizebox{0.99\linewidth}{!}{%
  \small
  \begin{tabular}{lccccccccc}
    \toprule
    \multirow{3}{*}{\vspace{-10pt}\textbf{Model name}} & \textbf{\multirowcell{3}{\\[-3pt]Model \\params.}} & \textbf{\multirowcell{3}{\\[-3pt]Beam\\size}} & \multicolumn{6}{c}{\textbf{CER [\%] ($\downarrow$)}} & \multirow{3}{*}{\vspace{-10pt}\textbf{RTF ($\downarrow$)}} \\
    \cmidrule(lr){4-9}
    & & & \textbf{\multirowcell{2}{\\[-6pt]Reazon \\test}} & \textbf{\multirowcell{2}{\\[-6pt]JSUT \\basic5000}} & \textbf{\multirowcell{2}{\\[-6pt]CV8.0 \\test}} & \multicolumn{3}{c}{\textbf{CSJ}} & \\
    \cmidrule(lr){7-9}
    & & & & & & \textbf{Eval1} & \textbf{Eval2} & \textbf{Eval3} & \\
    \midrule \midrule
    \multirow{3}{*}{reazonspeech-espnet-v1} & \multirow{3}{*}{90M}
      & 1  & 12.0 & 8.7  & 10.5 & 22.9 & 20.1 & 15.3 & \textbf{0.06} \\
    & & 5  & 9.1  & 7.6  & 9.0  & 21.4 & 18.6 & 15.3 & 0.20 \\
    & & 20 & 8.7  & 7.5  & 8.9  & 21.8 & 19.1 & 15.5 & 0.27 \\
    \midrule
    \multirow{2}{*}{Whisper-base} & \multirow{2}{*}{74M}
      & 1  & 38.6 & 23.5 & 25.7 & 29.9 & 27.5 & 25.9 & \textbf{0.06} \\
    & & 5  & 35.4 & 22.2 & 23.6 & 28.0 & 25.7 & 24.2 & 0.08 \\
    \midrule
    \multirow{2}{*}{Whisper-small} & \multirow{2}{*}{244M}
      & 1  & 27.3 & 14.1 & 14.9 & 21.7 & 21.5 & 18.0 & 0.10 \\
    & & 5  & 24.0 & 13.6 & 13.9 & 20.9 & 20.2 & 17.4 & 0.14 \\
    \midrule
    \multirow{2}{*}{Whisper-medium} & \multirow{2}{*}{769M}
      & 1  & 21.6 & 9.6  & 10.8 & 19.3 & 17.6 & 15.0 & 0.21 \\
    & & 5  & 20.6 & 9.4  & 10.7 & 18.5 & 17.3 & 14.9 & 0.28 \\
    \midrule
    \multirow{2}{*}{Whisper-large-v2} & \multirow{2}{*}{1,541M}
      & 1  & 21.8 & 8.1  & 9.4  & 23.7 & 17.1 & 19.6 & 0.31 \\
    & & 5  & 20.3 & 7.9  & 9.2  & 17.9 & 16.3 & 16.3 & 0.49 \\
    \midrule
    \multirow{2}{*}{Whisper-large-v3} & \multirow{2}{*}{1,541M}
      & 1  & 12.4 & \textbf{7.1}  & 8.2  & 17.2 & \textbf{15.6} & \textbf{14.8} & 0.30 \\
    & & 5  & 11.9 & \textbf{7.1}  & \textbf{8.0}  & \textbf{17.0} & \textbf{15.6} & 15.6 & 0.46 \\
    \midrule
    Proposed model\\
    \quad \textit{w/o DeepSpeed-Inference} & \multirow{2}{*}{3,708M} & 1 & \textbf{8.4} & 8.6 & 9.1 & 31.0 & 26.6 & 22.9 & 0.22 \\
    \quad \textit{w/ DeepSpeed-Inference} & & 1 & \textbf{8.4} & 8.6 & 9.2 & 30.9 & 26.6 & 22.9 & 0.15 \\
    \bottomrule
  \end{tabular}
  }
  \caption{
    ASR results of the proposed model and the publicly available ASR models on the ReazonSpeech test set, JSUT basic5000, CV8.0 test set, and three evaluation sets from CSJ.
    We compared different beam sizes in beam search decoding for inference of baseline models.
    Beam size 1 means greedy decoding.
  }
  \label{table:exp_bench}
\end{table*}

We evaluated the performance of the proposed model in comparison with publicly available ASR models.
The proposed model is denoted A3 in Table~\ref{table:exp_abl}.
We also compared the results of the proposed model with and without DeepSpeed-Inference optimization.
The following models were used as baseline E2E ASR models:
\begin{description}
  \item[reazonspeech-espnet-v1\footnotemark:]
  An encoder-decoder model using a Conformer-based encoder and Transformer-based decoder trained on the ReazonSpeech corpus~\citep{yin2023reazonspeech}.
  This is a character-level ASR model with joint decoding by combining both attention-based and CTC scores.
  Language model fused decoding was also performed using an external LSTM-based language model.
  \item[Whisper\footnotemark:]
  An encoder-decoder Transformer model trained on the large-scale weakly labeled speech data~\citep{radford2023robust}.
  The base, small, medium and large-v2 models were trained on 680k hours of multilingual speech data, including 7k hours of Japanese speech data.
  The large-v3 model was trained on 1M hours of weakly labeled and 4M hours of pseudo-labeled data.
\end{description}
\addtocounter{footnote}{-2}
\stepcounter{footnote}\footnotetext{\url{https://huggingface.co/reazon-research/reazonspeech-espnet-v1}}
\stepcounter{footnote}\footnotetext{\url{https://github.com/openai/whisper}}
These models operate on the 80-dimensional log Mel-spectrograms instead of raw waveforms, unlike the proposed model.

This experiment was performed using the ReazonSpeech test set, JSUT basic5000, and CV8.0 test sets and three evaluation sets from the Corpus of Spontaneous Japanese (CSJ)~\citep{maekawa2000spontaneous} to evaluate robustness to a broader domain.
The CSJ data are recordings of actual and mock conference lectures that contain more fillers, disfluency, and mispronounced words than the other corpora.
Note that the transcription tendency differs from that of the written text.
We measured the average real-time factor (RTF) on an NVIDIA T4 GPU using the first 100 utterances from JSUT basic5000.

The results are presented in Table~\ref{table:exp_bench}.
The inference speed of the proposed model improved 1.4 times faster with DeepSpeed-Inference while maintaining recognition accuracy.
Next, we focused on the results of recognition accuracy.
The results on ReazonSpeech test set shows that the proposed model outperforms the all beam size settings of reazonspeech-espnet-v1, which is the encoder-decoder model trained on the same ReazonSpeech corpus, in the in-domain scenario.
On the JSUT basic5000 and CV8.0 test sets, the proposed model achieved a lower CER than reazonspeech-espnet-v1 with a beam size of 1, although the CER was slightly worse than reazonspeech-espnet-v1 with beam sizes of 5 or 20.
The proposed model achieved comparable performance with a lower RTF and without complicated decoding procedures such as explicit language model fused decoding, indicating the high potential of decoder-only model-based ASR.
Compared with Whisper models, our model achieved the same performance as the large-v2 model, with a lower RTF than the medium version.
It does not reach the CER of the large-v3 model; one reason was the amount of training data used.
Training the proposed model on a larger speech corpus may close the gap.
Additional analysis of this experiment is provided in Appendix~\ref{sec:appendix_bench}.

All models exhibited worse CERs on CSJ Eval sets than on the other test sets.
This is because the compared models tended to ignore some words, such as fillers and disfluencies.
Note that our model performed significantly worse than the other models.
A possible reason for this difference is that in the proposed model, the weight of the LLM part was initialized with GPT pre-trained on written texts where fillers and disfluencies did not appear.
Since the proposed model leverages on the text generation ability of the LLM to recognize speech, it tends to drop some words corresponding to such disfluencies and fillers.

In addition, we found that the proposed model lost the first or last sentence when a given speech segment contained multiple sentences.
Since the CSJ corpus consists of 5 to 20 minutes of lecture speech and no manual annotation of full stops, the speech files were automatically segmented for evaluation, resulting in some segments containing multiple sentences.
This problem of dropping a part of the text to be recognized possibly stems from the architecture of the decoder-only Transformer.
In the standard encoder-decoder models, source-target attention layers can explicitly capture the alignment between speech and text, while in the proposed model, self-attention layers in the decoder-only Transformer capture the alignment.
This structural difference to capture alignments may affect the robustness of such a segment within multiple sentences.
This problem might be mitigated by increasing the diversity of the training data, such as by including longer speech segments.

\subsection{Domain adaptation}
\label{sec:exp_adapt}

Adapting models to specific domains is widely used to address the issue of robustness to out-of-domain data, including speech recognition.
However, since the amount of adaptive data is often limited in practical use, it is beneficial to consider efficient adaptation for proposed models with a large number of model parameters.
Therefore, we conducted additional experiments to evaluate the domain adaptation performance of the proposed model with PEFT, which has recently been demonstrated to have significant adaptation performance for various generation tasks such as LLM and image generation.
The base model without adaptation is A3 in Table~\ref{table:exp_abl}.
Considering with the practical convenience required for domain adaptation, this section explores PEFT using LoRA for both HuBERT and GPT.
We set the LoRA parameter to $R=32$ to adjust the self-attention parameters for HuBERT and GPT.
We used the same objective~\eqref{eq:llm_objective} as for regular training on a single A100 80GB GPU with a batch size of 64.
The total number of updating parameters in adapting all modules, including HuBERT and GPT, was only 22 million.
We performed domain adaptation experiments with five different scenarios using two dataset: CV8.0 train set and CSJ train set.
Two data sizes were used as adaptation scenarios for CSJ: core data only (34 hours) and all data (520 hours).
The case of simultaneous adaptation to CV8.0 and CSJ data was also considered.

The results are listed in Table~\ref{table:exp_adapt}.
The models adapted to CV8.0, slightly improved the CER on CV8.0 test set.
This limited improvement could be attributed to the small amount of adaptation data available.
It should also be noted that the CV8.0 adapted models also outperformed the non-adapted model not only on the CV8.0 dataset but also on the JSUT and CSJ datasets.
The ReazonSpeech corpus, which we used as training data for the base model, suffers from a partial mismatch between speech and text, owing to the automatic collection and segmentation of TV program recordings with a pre-trained ASR model.
Although the CV8.0 corpus contains speeches of poor recording quality and by non-native speakers, it has been verified by comparing transcriptions to recorded speech by the contributors.
This result suggests that model optimization with limited human-validated data improves recognition accuracy, regardless of the specific domain.

\begin{table}
  \centering
  \resizebox{0.99\linewidth}{!}{%
    \begin{tabular}[t]{cccccccc}
      \toprule
      \multicolumn{3}{c}{\textbf{Adaptation}} & \multicolumn{5}{c}{\textbf{CER [\%] ($\downarrow$)}} \\
      \cmidrule(lr){1-3} \cmidrule(lr){4-8}
      \textbf{\multirowcell{2}{\\[-10pt]HuBERT}} & \textbf{\multirowcell{2}{\\[-10pt]Bridge\\network}} & \textbf{\multirowcell{2}{\\[-10pt]GPT}} & \hspace{-3mm}\textbf{\multirowcell{2}{\\[-10pt]JSUT\\basic5000}}\hspace{-3mm} & \textbf{\multirowcell{2}{\\[-10pt]CV8.0\\test}} & \multicolumn{3}{c}{\textbf{CSJ}} \\
      \cmidrule(lr){6-8}
      & & & & & \textbf{Eval1} & \textbf{Eval2} & \textbf{Eval3} \\
      \midrule\midrule
      \multicolumn{7}{l}{\textbf{No adaptation}} \\
      \quad      &            &            & 8.6 & 9.2 & 30.9 & 26.6 & 22.9 \\
      \midrule
      \multicolumn{7}{l}{\textbf{Adaptation to CV8.0 train (9 hours)}} \\
                 &            & \checkmark & 7.6 & \greycell \textbf{8.6} & 28.1 & 24.9 & 20.7 \\
                 & \checkmark & \checkmark & 7.7 & \greycell 8.9 & 30.5 & 26.2 & 23.0 \\
      \checkmark & \checkmark & \checkmark & 7.7 & \greycell 8.9 & 29.4 & 25.8 & 22.4 \\
      \midrule
      \multicolumn{7}{l}{\textbf{Adaptation to CSJ core (34 hours)}} \\
                 &            & \checkmark & 8.7 & 11.2 & \greycell 7.9 & \greycell 7.0 & \greycell 4.9 \\
                 & \checkmark & \checkmark & 8.8 & 15.1 & \greycell 7.4 & \greycell 6.2 & \greycell 4.4 \\
      \checkmark & \checkmark & \checkmark & 8.5 & 12.4 & \greycell 7.0 & \greycell 6.2 & \greycell 4.1 \\
      \midrule
      \multicolumn{7}{l}{\textbf{Adaptation to CSJ all (520 hours)}} \\
                 &            & \checkmark & 8.6 & 14.9 & \greycell 6.6 & \greycell 5.3 & \greycell 4.1 \\
                 & \checkmark & \checkmark & 9.5 & 15.9 & \greycell 5.8 & \greycell 4.6 & \greycell \textbf{3.4} \\
      \checkmark & \checkmark & \checkmark & 8.7 & 15.5 & \greycell \textbf{5.4} & \greycell \textbf{4.4} & \greycell \textbf{3.4} \\
      \midrule
      \multicolumn{7}{l}{\textbf{Adaptation to CV8.0 train and CSJ core (43 hours)}} \\
                 &            & \checkmark & \textbf{7.2} & \greycell 8.9 & \greycell 7.9 & \greycell 6.9 & \greycell 4.7 \\
                 & \checkmark & \checkmark & 8.1 & \greycell 8.7 & \greycell 7.5 & \greycell 6.3 & \greycell 4.4 \\
      \checkmark & \checkmark & \checkmark & 7.7 & \greycell 8.7 & \greycell 7.2 & \greycell 6.1 & \greycell 4.3 \\
      \midrule
      \multicolumn{7}{l}{\textbf{Adaptation to CV8.0 train and CSJ all (529 hours)}} \\
                 &            & \checkmark & 7.8 & \greycell 9.0 & \greycell 6.6 & \greycell 5.3 & \greycell 4.0 \\
                 & \checkmark & \checkmark & 7.9 & \greycell 9.0 & \greycell 5.6 & \greycell 4.5 & \greycell 3.8 \\
      \checkmark & \checkmark & \checkmark & 7.8 & \greycell 9.1 & \greycell \textbf{5.4} & \greycell 4.5 & \greycell 3.5 \\
      \bottomrule
    \end{tabular}
  }
  \caption{
    ASR results with different adaptation settings.
    The gray color means the same domain of test sets as that of adaptation data.
  }
  \label{table:exp_adapt}
\end{table}

The models adapted to CSJ significantly improved CER on CSJ evaluation sets in both the core and all data settings.
However, the adapted models had a worse CER for out-of-domain test data such as JSUT basic5000 and CV8.0, in contrast to the results for the CV8.0 adaptation.
Upon inspecting the recognition results of the adapted model, we noticed that fillers that should only be present in CSJ data were sometimes included in the recognized texts of JSUT basic5000 and CV8.0, even though they were not present in the actual speech.
The possible reason for this is that model optimization could have been promoted to insert such fillers as a text-generation task rather than an ASR task.
This hallucination phenomenon was particularly prominent in the CV8.0 test set.
The CER on CV8.0 test of the adapted models was significantly lower than that of JSUT basic5000 compared with the non-adapted models.
The features of CV8.0 speech are similar to CSJ with some noise and reverberation, whereas JSUT consists of clean speech.
Similarities in the characteristics between CV8.0 and CSJ datasets caused the CER on CV8.0 to worsen the more the model adapted to the CSJ dataset.

Models adapted to both CV8.0 and CSJ enabled the adaptation of CSJ while maintaining the same or better recognition performance for CV8.0 test set.
Furthermore, these adapted models were able to achieve a better CER for JSUT basic5000 as well as models adapted to only CV8.0.

Appendix~\ref{sec:appendix_adapt} also discusses the results of independently adapting each model within the proposed framework.
As an additional example of extending the proposed model, we conducted adaptation experiments to output a Japanese mora sequence instead of plain text as ASR outputs in Appendix~\ref{sec:appendix_mora}.

It is difficult to adapt to the target domain while maintaining the recognition accuracy for the out-of-target domain of adaptation, depending on the particularities of the adaptation domain.
Nevertheless, owing to the recent PEFT methods, because fine-tuning a large-scale model does not always require enormous computational costs or large amounts of adaptation data, domain adaptation each time for each domain is not a significant problem for practical application.

\section{Conclusions}

This paper proposed a fully E2E ASR model that integrates a pre-trained speech representation model and an LLM.
In the proposed model, the pre-trained HuBERT and GPT are connected by a convolution-based bridge network and are fully fine-tuned, where the bridge network passes meaningful continuous latent representations extracted from the speech waveform sample to the LLM as speech prompts.
Experiments demonstrated that the proposed model achieved a performance comparable to that of publicly available modern ASR models.
We also investigated the domain adaptation capabilities of different text and speech domains using PEFT.
The strength of the proposed model lies in its potential to be easily integrated with various speech and language pre-trained models, including multilingual LLMs such as~\citet{yong2023bloom}.
Future work includes further comparison of existing E2E baseline models, use of different pre-training models, and extension to multilingual ASR.

While this paper focused on the ASR task as an extension of pre-trained models, we believe that it is essential to investigate ways to improve performance in specific domains to advance multi-task speech-text processing capabilities by leveraging these models.
Extending pre-trained models to encompass a wider range of multimodal speech and text tasks, such as automatic speech translation, speech question answering, and other related areas, is a promising future direction.

\section{Limitations}

Based on our experimental results and analyses, we discuss some limitations and avenues for addressing them.

The primary consideration in the proposed method is the amount of data required.
As mentioned in Section~\ref{sec:exp_setup} and Appendix~\ref{sec:appendix_data}, the proposed model was trained on over 15,000 hours of speech/text pair data to integrate pre-trained HuBERT and GPT.
This is a concern when considering extensions to languages for which such large amounts of speech data do not exist, especially for low-resource languages.
Although it is intuitively expected that the performance of the integrated model will decrease as the amount of data decreases, an investigation into the actual trade-off between data volume and performance is warranted.
Future directions to overcome this limitation include exploring data augmentation and applying transfer learning techniques from other languages.
The adaptation experiments presented in Section~\ref{sec:exp_adapt} are expected to facilitate extension of multilingual ASR and transfer learning to low-resource languages.

Moreover, given that it is impractical to construct pre-trained models exclusively with low-resource languages, one direction is to use a multilingual pre-training model~\citep{yong2023bloom}.
However, since the parameter size tends to increase with the size of the training data, the relationship between the model size and pre-training model size should be investigated.

The second limitation concerns inference speed.
A significant challenge of the proposed method is the difficulty of performing inferences on low-resource devices, mainly because of the large number of parameters in the LLM.
Recently, ASR on compact devices such as smartphones and smart speakers has become widespread as a input method for natural human-computer interactions.
In these scenarios, there is demand for on-device ASR from a privacy perspective and off-line inference.

These limitations are not unique to the proposed model but are a common issue in recent multimodal processing that utilizes self-supervised learning models and LLMs.
Starting various speech and language processing tasks on the common foundation of these pre-trained models leads to a multifaceted understanding of these models.
Research addressing these challenges is expected to continue.

The third limitation relates to the evaluation method for ASR systems in languages that exhibit orthographic variations.
For example, in the case of Japanese, the correct forms are written in \emph{hiragana} and \emph{kanji}.
It should be noted that the current standard evaluation metric in ASR, which directly uses the Levenshtein distance against the correct text, fails to proper consider orthographic variations.

\section{Ethics Statement}

The proposed model shares ethical concerns with the other ASR models.
Because it uses human speech as input, differences in user attributes, such as age, gender, dialect, and whether the speaker is native or non-native, could potentially impact recognition accuracy.
For instance, it has been reported that various characteristics of elderly voices differ significantly from those of younger people, such as hoarseness, slower speech, and unclear pronunciation~\citep{winkler2003aging,miyazaki2010acoustic}.
In addition, non-native speakers may produce more ambiguous pronunciations close to their native language.
From a text perspective, it would also be ethically controversial if the recognition results showed extreme accuracy or inaccuracy only for specific topics, such as certain political or religious beliefs.
Given the data-driven nature of recent ASR approaches, the variability in the domains of speech and text data used for training requires careful consideration in the practical application of ASR systems.

This work used publicly available datasets and pre-trained models, which ensured the transparency of the experiments.
The additional data used in the evaluation and adaptation experiments are also publicly available and widely used in Japanese speech-related research.

The proposed model output the recognized text by a decoder-only Transformer as a next-token prediction, similar to recent LLMs.
This may cause hallucinations, as is the case with existing LLMs, leading to the generation of a recognized text that differs from the actual spoken content.
However, such recognized errors are a potential risk common to all speech recognition models, not just the proposed method, and further improvements are required from both data and method perspectives.

\bibliography{mybib}

\newpage
\appendix

\section{Dataset and Preprocessing Details}
\label{sec:appendix_data}

We used the following datasets in the experiments: ReazonSpeech, JSUT, Common Voice 8.0, and CSJ.
All datasets are publicly available and used in the Japanese speech-related research community.
The details of each dataset are summarized below.

\paragraph{ReazonSpeech:}
This corpus is a large-scale Japanese speech corpus constructed from audio and caption data from recorded TV programs~\citep{yin2023reazonspeech}.
It encompasses various sounds, ranging from clean audio to recordings mixed with audience cheers and background music.

We used 1,000 utterances as development data, 1,000 different utterances as test data, and the remainder as training data by following a publicly available ASR recipe\footnote{\url{https://github.com/espnet/espnet/tree/105e532/egs2/reazonspeech/asr1}}.
We found some duplication between the utterances in the development/test set and those in the training data.
Therefore, we performed a transcript-level de-duplication to development and test set and finally removed utterances that matched more than 5 characters.
This corpus contains audio files with little or no human speech because of the properties of the corpus, which is automatically constructed from audio and caption data from recorded TV programs.
Thus, we pruned the data with a few speech segments based on the length of the transcribed text and speech waveforms, wherein 10 million utterances (approximately 15,700 hours) were used.

\paragraph{JSUT:}
This corpus is a Japanese speech corpus designed for end-to-end speech synthesis~\citep{sonobe2017jsut}.
It consists of clean speech data recorded with a 48 kHz sampling in an anechoic room by one female native Japanese speaker.

This corpus includes nine subsets, and basic5000 subset was used for the evaluation of the ASR models.
This subset contains 5,000 utterances, covering all main pronunciations of daily-use characters and their individual onyomi (Japanese readings) and kunyomi (Chinese readings).
The audio was downsampled to 16 kHz for experiments.

\begin{figure*}[t]
  \centering
  \subfloat[ReazonSpeech development set \label{fig:appendix_edit1:valid}]{
    \includegraphics[width=0.45\linewidth]{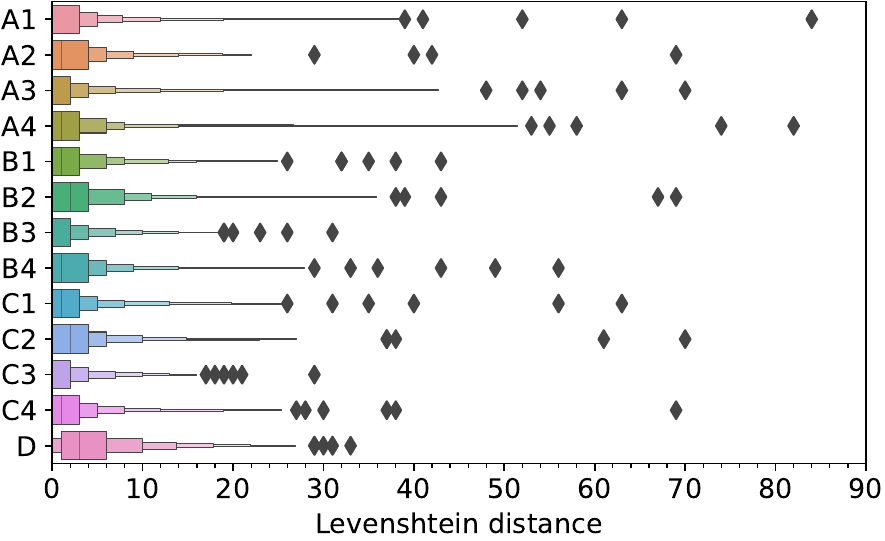}
  }
  \hspace{5mm}
  \subfloat[ReazonSpeech test set \label{fig:appendix_edit1:test}]{
    \includegraphics[width=0.45\linewidth]{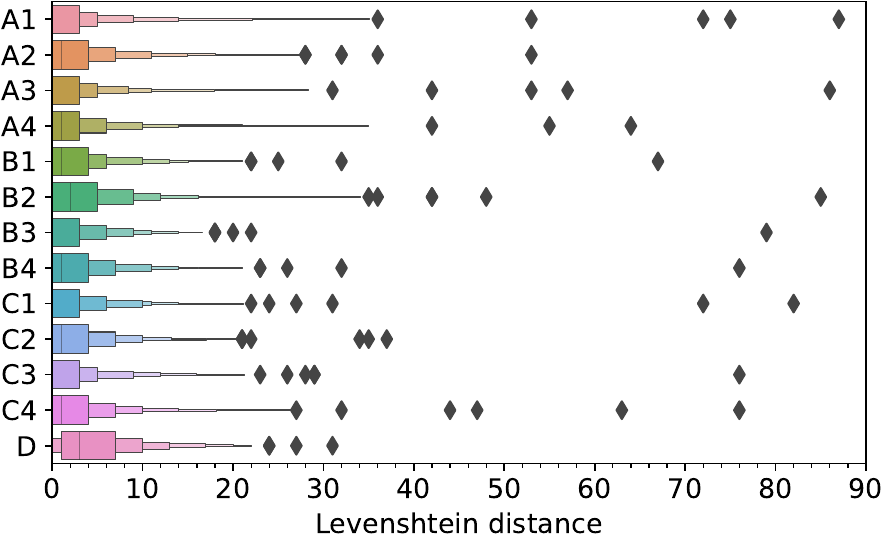}
  }
  \\
  \subfloat[JSUT basic5000 \label{fig:appendix_edit1:jsut}]{
    \includegraphics[width=0.45\linewidth]{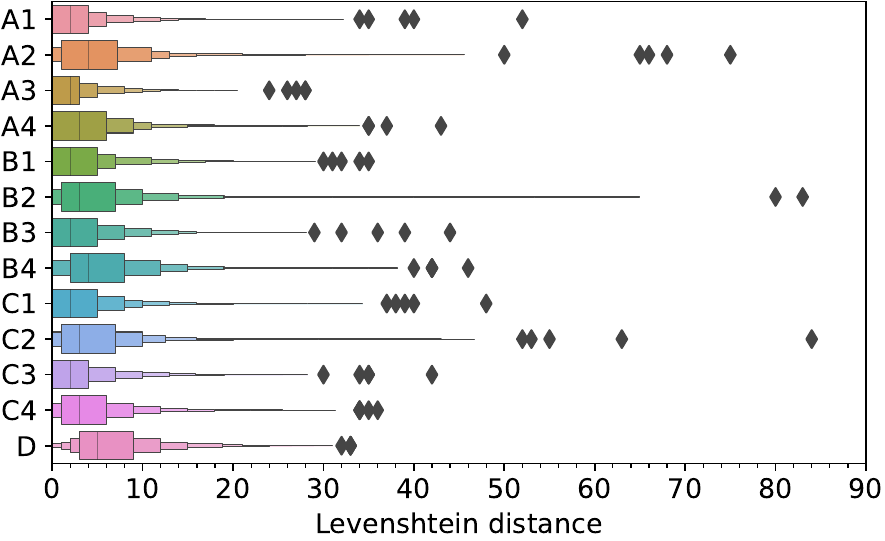}
  }
  \hspace{5mm}
  \subfloat[CV8.0 test set \label{fig:appendix_edit1:cv8}]{
    \includegraphics[width=0.45\linewidth]{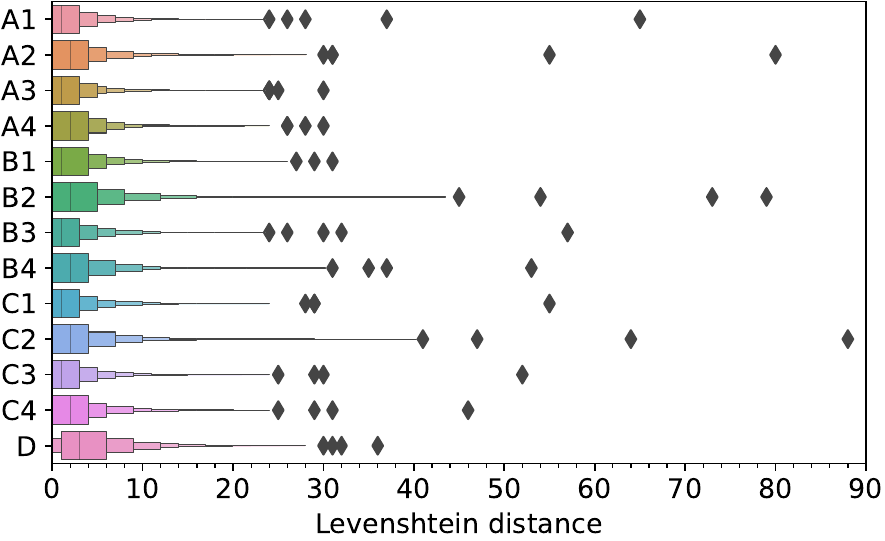}
  }
  \caption{
    Letter-value plot of Levenshtein distance for ablation study of the proposed model.
    The IDs of each model are the same as the IDs in the Table~\ref{table:exp_abl}.
    }
  \label{fig:appendix_edit1}
  \vspace{-1mm}
\end{figure*}

\paragraph{Common Voice:}
This corpus is a massively multilingual, multispeaker dataset of transcribed speech intended to support research and development in speech technologies, particularly for ASR~\citep{ardila2020common}.
Because it relies on crowdsourcing for data collection and validation, there are some recordings with a lot of background noise due to poor recording environments and recordings from non-native speakers.

For our study, we used the Japanese subset of Common Voice 8.0.
The test set was employed to evaluate the ASR models, and the training and development sets were further used for domain adaptation experiments detailed in Section~\ref{sec:exp_adapt} and Appendix~\ref{sec:appendix_adapt}.
Since the audio clips are provided as 16-bit MPEG-3 files with a 48 kHz sampling rate, they were downsampled to 16 kHz for experiments.

\paragraph{CSJ:}
This corpus contains 660-hour spontaneously uttered Japanese speech and the morphologically annotated transcriptions.
Most of this corpus consists of data recorded from academic presentation speech and simulated public speaking.
Since the length of such speech is mostly 10 to 25 minutes, the audio files were segmented based on the annotated pause information, following a publicly available ASR recipe\footnote{\url{https://github.com/kaldi-asr/kaldi/tree/master/egs/csj/s5/local/csj_make_trans}}.
This corpus includes speeches with text domains not commonly found in standard speech corpora, such as those from science and engineering, humanities, and society-related presentations.
It also transcribes fillers, disfluency, and mispronounced words found in spontaneous speech, which significantly differs from the ReazonSpeech corpus.
Hence, this dataset was utilized to evaluate the out-of-domain performance of the ASR models in Section~\ref{sec:exp_bench} and as adaptation data for domain adaptation in Section~\ref{sec:exp_adapt} and Appendix~\ref{sec:appendix_adapt}.
This corpus contained the pronunciative form notation, which was used in the additional experiments in Appendix~\ref{sec:appendix_mora}.

The sampling rate of provided audio samples was 16 kHz.
The CSJ files are grouped into Core, which contains detailed annotations but limited audio data, and others, which do not.
The same three evaluation sets presented in the CSJ corpus were used as test data.

\begin{table*}[tp]
  \centering
  \resizebox{0.99\linewidth}{!}{
    \begin{tabular}[t]{ccccccccccccccccccc}
      \toprule
      & \multicolumn{2}{c}{\textbf{Model description}} & \multicolumn{16}{c}{\textbf{CER [\%] ($\downarrow$)}} \\
      \cmidrule(lr){2-3} \cmidrule(lr){4-19}
      \textbf{IDs} & \textbf{\multirowcell{2}{\\[-10pt]HuBERT}} & \textbf{\multirowcell{2}{\\[-10pt]GPT}} & \multicolumn{4}{c}{\textbf{ReazonSpeech dev}} & \multicolumn{4}{c}{\textbf{ReazonSpeech test}} & \multicolumn{4}{c}{\textbf{JSUT basic5000}} & \multicolumn{4}{c}{\textbf{CV8.0 test}} \\
      \cmidrule(lr){4-7} \cmidrule(lr){8-11} \cmidrule(lr){12-15} \cmidrule(lr){16-19}
      & & & \textbf{short} & \textbf{mid} & \textbf{long} & \textbf{all} & \textbf{short} & \textbf{mid} & \textbf{long} & \textbf{all} & \textbf{short} & \textbf{mid} & \textbf{long}* & \textbf{all} & \textbf{short} & \textbf{mid} & \textbf{long} & \textbf{all} \\
      \midrule\midrule
      \multicolumn{6}{l}{\hspace{-0.6em} \textbf{HuBERT + GPT (w/ Downsampling)}} \\
      A1 & Frozen  & Full FT &  8.1 &  6.1 & 24.2 &  8.9 &  9.7 &  7.1 & 21.9 &  9.7 & 10.4 & 10.4 &  7.1 & 10.4 &  9.7 & 10.6 & -- & 10.2 \\
      A2 & Full FT & Frozen  & 10.4 &  9.6 & 15.0 & 10.5 & 14.6 &  9.7 & 10.9 & 10.8 & 19.7 & 18.3 & 31.0 & 19.1 & 11.7 & 14.7 & -- & 13.4 \\
      A3 & Full FT & Full FT &  7.6 & \bf{5.4} & 23.0 &  8.2 & 9.0 & \bf{6.2} & 18.0 &  8.4 & \bf{8.6} & \bf{8.6} &  5.8 & \bf{8.6} & \bf{8.9} & \bf{9.4} & -- & \bf{9.2} \\
      A4 & Full FT & PEFT    &  8.8 &  8.1 & 28.3 & 10.9 & 10.3 &  8.0 & 21.7 & 10.4 & 15.0 & 14.5 & 12.3 & 14.8 & 11.3 & 12.7 & -- & 12.1 \\
      \midrule
      \multicolumn{6}{l}{\hspace{-0.6em} \textbf{HuBERT-CTC + GPT (w/ CTC remove)}} \\
      B1 & Frozen  & Full FT &  9.0 &  8.1 & 11.0 &  8.7 & 12.2 &  7.9 & 10.1 &  9.1 & 12.0 & 12.2 &  9.0 & 12.1 & 10.6 & 12.8 & -- & 11.9 \\
      B2 & Full FT & Frozen  & 12.8 & 13.0 & 14.7 & 13.2 & 15.8 & 12.5 & 15.4 & 13.6 & 14.8 & 22.6 & 25.8 & 18.5 & 14.6 & 21.2 & -- & 18.4 \\
      B3 & Full FT & Full FT & \bf{7.1} &  5.8 &  8.0 & \bf{6.4} &  9.9 &  7.0 & \bf{8.8} & \bf{7.8} & 12.8 & 11.2 &  9.0 & 12.1 &  9.8 & 10.9 & -- & 10.4 \\
      B4 & Full FT & PEFT    & 10.9 &  8.6 & 14.2 &  9.9 & 12.0 &  9.3 & 11.8 & 10.2 & 23.2 & 19.7 & 20.6 & 21.6 & 12.5 & 16.1 & -- & 14.6 \\
      \midrule
      \multicolumn{6}{l}{\hspace{-0.6em} \textbf{HuBERT-CTC + GPT (w/ CTC average)}} \\
      C1 & Frozen  & Full FT &  9.0 &  6.9 & 15.0 &  8.5 & 10.9 &  7.4 & 11.2 &  8.6 & 12.4 & 11.4 & 11.6 & 11.9 &  9.9 & 11.0 & -- & 10.6 \\
      C2 & Full FT & Frozen  & 12.1 &  9.8 & 14.9 & 11.0 & 12.0 & 10.1 &  9.9 & 10.5 & 15.7 & 18.5 & 15.5 & 17.0 & 12.4 & 17.1 & -- & 15.1 \\
      C3 & Full FT & Full FT &  7.5 &  6.1 & \bf{6.8} &  6.6 & \bf{8.8} &  7.1 &  9.4 & \bf{7.8} & 12.3 & 10.6 & \bf{3.2} & 11.5 &  9.2 & 10.0 & -- &  9.7 \\
      C4 & Full FT & PEFT    &  9.2 &  8.0 & 12.8 &  8.9 & 11.7 &  9.0 & 12.9 & 10.1 & 15.2 & 15.5 & 11.6 & 15.3 & 11.1 & 14.4 & -- & 13.0 \\
      \midrule
      \multicolumn{6}{l}{\hspace{-0.6em} \textbf{HuBERT-CTC}} \\
      D  & Frozen & - & 16.9 & 16.8 & 18.3 & 16.9 & 15.9 & 16.7 & 16.6 & 16.3 & 24.1 & 23.0 & 26.3 & 23.8 & 21.3 & 21.7 & -- & 21.5 \\
      \bottomrule
    \end{tabular}
  }
  \caption{
    Detailed ASR results of the ablation study on the ReazonSpeech development and test set, JSUT basic5000, and CV8.0 test set.
    The CER was calculated based on the groups categorizing audio length as ``short,'' ``mid,'' and ``long.''
    The number of samples in each group is listed in Table.~\ref{table:appendix_eval_subset}.
    Note that the results for long JSUT basic5000 (marked as ``*'') are unreliable because they were computed from only two long utterance samples.
  }
  \label{table:appendix_abl}
\end{table*}

\begin{table}[t]
  \centering
  \resizebox{0.99\linewidth}{!}{
    \begin{tabular}[t]{lcccc}
      \toprule
      \multirowcell{2}{\\[-10pt]\textbf{Dataset}} & \textbf{Short} & \textbf{Mid} & \textbf{Long} & \multirowcell{2}{\\[-10pt]\textbf{Total}} \\
      & \textbf{($<$5.1s)} & \hspace{-1mm}\textbf{(5.1s -- 15.9s)}\hspace{-1mm} & \textbf{($\geq$15.9s)} & \\
      \midrule\midrule
      ReazonSpeech dev   & 332   & 349   & 28 & 709 \\
      ReazonSpeech test  & 292   & 375   & 31 & 698 \\
      JSUT basic5000     & 3,498 & 1,500 & 2  & 5,000 \\
      CV8.0 test         & 2,599 & 1,884 & 0  & 4,483 \\
      \bottomrule
    \end{tabular}
  }
  \caption{
    Number of short, medium, and long utterances in the development and test data.
    The boundaries for each group, at 5.1 seconds and 15.9 seconds, correspond to the median and maximum utterance lengths in the filtered ReazonSpeech training set used to train the proposed model.
  }
  \label{table:appendix_eval_subset}
\end{table}

\section{Supplementary Experimental Analysis}

\subsection{Analysis of ablation study}
\label{sec:appendix_abl}

We analyzed the ablation study for the proposed model in Section~\ref{sec:exp_abl}.
Figure~\ref{fig:appendix_edit1} shows the letter-value plot of the Levenshtein distances between the target and recognized texts of each utterance on the ReazonSpeech development and test set, the JSUT basic5000, and the CV8.0 test set.

The HuBERT-CTC model (D) had larger distances on average than other models integrating GPT, yet it had fewer outliers across each test set.
In contrast, the overall tendency of the models integrating HuBERT and GPT was to have a wider range of Levenshtein distance values and more outliers, especially for the ReazonSpeech development and test sets.
This indicates that the integrated model is less stable in sequence generation than the HuBERT-CTC model due to autoregressive decoding by the GPT.

For a more detailed analysis, each development and test set was divided into three groups based on the length of the speech: short, mid, and long, and the CER was calculated for each group.
The results of the CER aggregated for each group are summarized in Table~\ref{table:appendix_abl}, and the number of utterances for each group is summarized in Table~\ref{table:appendix_eval_subset}.

Table~\ref{table:appendix_abl} shows that the models with downsampling-based sequence compression (A1--A4) heavily depend on the length of the speech for their performance.
In particular, the GPT fine-tuned models (A1, A3, and A4) suffer a significant degradation in CER when processing long utterances that do not appear in the training data.
This is because downsampling, with its fixed compression rate, is sensitive to temporal structure of speech, such as speaking rate and the number and frequency of pauses.
Conversely, the models with CTC remove (B1--B4) and CTC average (C1--C4) are less sensitive to utterance length than those with downsampling-based ones.
In fact, the HuBERT-CTC model (D4) is likewise hardly affected by speech utterance length.
Thus, it can be seen that CTC-based sequence compression is superior in terms of robustness to utterance length.

\begin{figure*}[t]
  \centering
  \subfloat[JSUT basic5000 \label{fig:appendix_edit2:jsut}]{
    \includegraphics[width=0.47\linewidth]{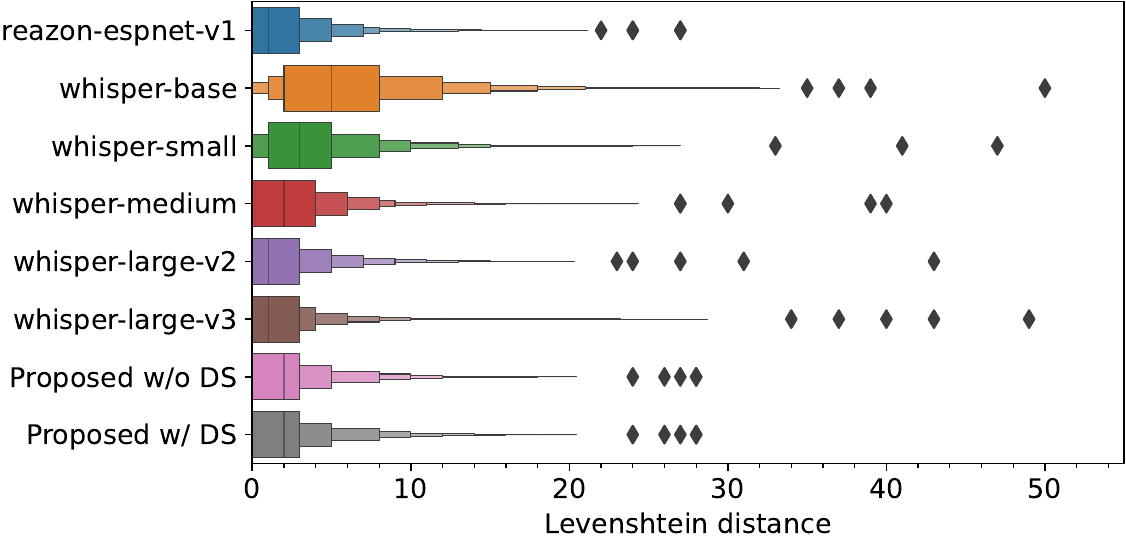}
  }
  \hspace{2mm}
  \subfloat[CV8.0 test set \label{fig:appendix_edit2:cv8}]{
    \includegraphics[width=0.47\linewidth]{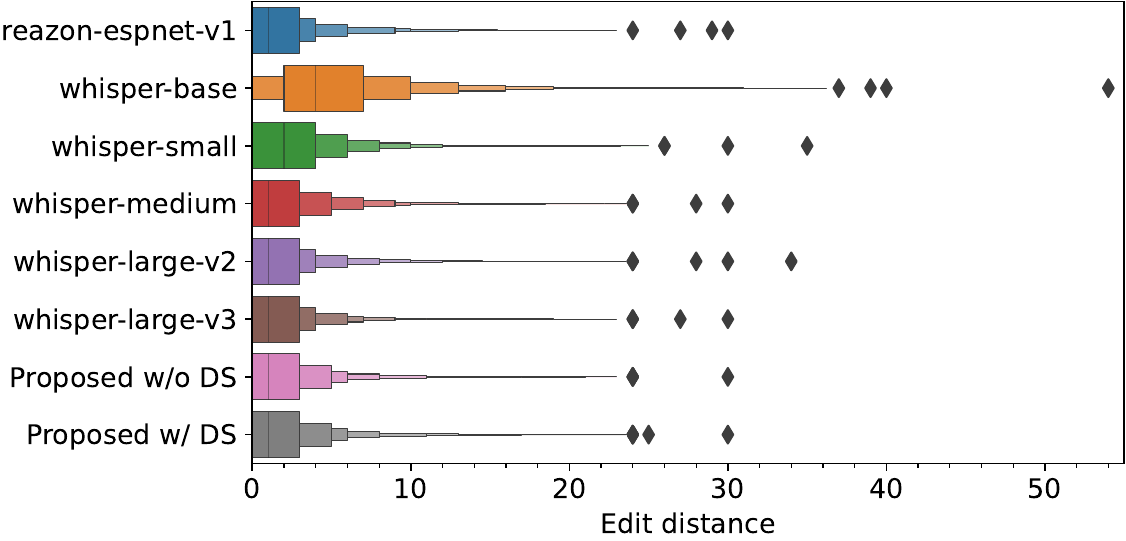}
  }
  \caption{
    Letter-value plot of Levenshtein distance for the comparison of the proposed model with the publicly available ASR models.
    The proposed model employed greedy search, while the comparative models used beam search with a beam size of 5.
    ``DS'' denotes DeepSpeed-Inference.
  }
  \label{fig:appendix_edit2}
  \vspace{-1mm}
\end{figure*}

\subsection{Analysis of comparison with other ASR models}
\label{sec:appendix_bench}

We conducted additional analysis on the comparative experiment results between our proposed model and publicly available models in Section~\ref{sec:exp_bench}.
Figure~\ref{fig:appendix_edit2} shows the letter-value plot of the Levenshtein distances between the target and recognized texts of each utterance on the JSUT basic5000 and the CV8.0 test set.

Each Whisper model, particularly with the JSUT basic5000 dataset, resulted in a higher incidence of undesirable outliers.
Upon examining these recognition results, it became apparent that the errors were more frequently due to orthographic variations (such as \emph{kanji} in Japanese texts being transcribed as \emph{hiragana} or \emph{katakana}) rather than failures inherent to autoregressive models (such as repeating some words).
These orthographic variations are not incorrect but somewhat unnatural as Japanese text.
The reason why Whisper struggles with such issues may stem from the quality of the transcribed texts derived from training data automatically collected from the Internet.

Meanwhile, as with reazonspeech-espnet-v1, the proposed model did not generate samples with a huge Levenshtein distance for either test set, unlike Whisper model family.
This demonstrates that integrating pre-trained speech and language models enables the construction of an ASR model whose performance is comparable to that of the typical source-target attention-based ASR model utilizing external language models in practical applications.

\section{Additional Investigation for the Proposed Model}
\subsection{Decoding strategies}
\label{sec:appendix_decode}

In this section, we compare the decoding strategies for the proposed model.
We conducted experiments using four different decoding strategies: greedy search, beam search, top-$k$ sampling~\citep{fan2018hierarchical}, and nucleus sampling~\citep{holtzman2020curious}, with different parameters.

\begin{table}[t]
  \centering
  \resizebox{0.95\linewidth}{!}{
    \begin{tabular}[t]{lcc}
      \toprule
      \multirow{2}{*}{\vspace{-7pt}\textbf{Decoding strategies}} & \multicolumn{2}{c}{\textbf{CER [\%] ($\downarrow$)}} \\
      \cmidrule(lr){2-3}
      & \textbf{Reazon dev} & \textbf{Reazon test}  \\
      \midrule\midrule
      Greedy & \textbf{8.2} & \textbf{8.4} \\
      \midrule
      Beam search (size=2) & 8.6 & 8.9 \\
      Beam search (size=3) & 8.6 & 8.7 \\
      Beam search (size=4) & 8.4 & 8.7 \\
      Beam search (size=5) & 8.7 & 8.6 \\
      \midrule
      Top-$k$ sampling ($k$=2) & 9.3 & 9.3 \\
      Top-$k$ sampling ($k$=3) & 9.9 & 9.6 \\
      Top-$k$ sampling ($k$=4) & 9.9 & 9.9 \\
      Top-$k$ sampling ($k$=5) & 10.5 & 10.3 \\
      \midrule
      Nucleus sampling ($p$=0.1) & \textbf{8.2} & \textbf{8.4} \\
      Nucleus sampling ($p$=0.3) & 8.3 & 8.5 \\
      Nucleus sampling ($p$=0.5) & 8.4 & 8.7 \\
      Nucleus sampling ($p$=0.7) & 8.9 & 8.8 \\
      Nucleus sampling ($p$=0.9) & 9.6 & 9.3 \\
      \bottomrule
    \end{tabular}
  }
  \caption{Comparison of decoding strategies for the proposed model.}
  \label{table:appendix_decode}
\end{table}

The results are listed in Table~\ref{table:appendix_decode}.
Table~\ref{table:appendix_decode} shows that introducing beam search did not result in a noticeable improvement in CER, unlike typical encoder-decoder ASR models such as reazonspeech-espnet-v1 and Whisper model family, as shown in Table~\ref{table:exp_bench}.
This could be because the proposed model with the LLM has more parameters and a higher modeling capability than other ASR models, allowing the LLM to decode the text tokens of the recognition results from the given speech prompts with high confidence.
On the other hand, unlike popular text generation tasks by LLMs, such as story generation~\citep{holtzman2020curious}, Table~\ref{table:appendix_decode} also shows that decoding strategies involving sampling are ineffective for ASR tasks.
The speech prompt is a powerful condition in ASR tasks, making it sufficient to perform likelihood-based deterministic decoding.

\subsection{Integrating other speech pre-trained models}
\label{sec:appendix_encoder}

We compared wav2vec~2.0 Base~\citep{baevski2020wav2vec2} and the encoder of Whisper small~\citep{radford2023robust} with HuBERT Base~\citep{hsu2021hubert} as pre-trained models for initializing the speech encoder.
All models contain 12 transformer blocks, model dimension 768, inner feed-forward network dimension 3,072, and 12 attention heads.
wav2vec~2.0 takes speech waveform directly as input and then is fed into a convolutional waveform encoder as the same in HuBERT, while Whisper encoder takes an 80-dimensional mel-spectrogram computed with a hop size of 10 ms.
We used a pre-trained japanese-wav2vec2-base\footnote{\url{https://huggingface.co/rinna/japanese-wav2vec2-base}} model~\citep{sawada2024release}, which was a wav2vec~2.0 Base model trained on the ReazonSpeech corpus.
Under the condition in GPT with full-parameter fine-tuned, we compared whether or not the Speech encoder is full-parameter fine-tuned in the case of each model.

\begin{table}[t]
  \centering
  \resizebox{0.95\linewidth}{!}{
    \begin{tabular}[t]{ccccc}
      \toprule
      \multicolumn{2}{c}{\textbf{\multirowcell{2}{\\[-10pt]Architecture\\of speech encoder}}} & \multicolumn{3}{c}{\textbf{CER [\%] ($\downarrow$)}} \\
      \cmidrule(lr){3-5}
      & & \textbf{Reazon} & \textbf{JSUT} & \textbf{CV8.0} \\
      \midrule\midrule
      \multirow{2}{*}{wav2vec2.0}
      & Frozen  & 31.1 & 26.9 & 37.3 \\
      & Updated & 8.5 & 10.4 & 10.8 \\
      \midrule
      \multirow{2}{*}{HuBERT}
      & Frozen  & 9.7  & 10.4 & 10.2 \\
      & Updated & 8.4 & \textbf{8.6} & \textbf{9.2} \\
      \midrule
      \multirow{2}{*}{Whisper-small}
      & Frozen  & 9.7 & 10.1 & 9.8 \\
      & Updated & \textbf{8.2} & \textbf{8.6} & 10.0 \\
      \bottomrule
    \end{tabular}
  }
  \caption{ASR results with different speech encoders.}
  \label{table:appendix_encoder}
\end{table}

The ASR results using the ReazonSpeech test set, JSUT basic5000, and CV8.0 test set are listed in Table~\ref{table:appendix_encoder}.
HuBERT integrated models demonstrate a lower CER compared to wav2vec 2.0 integrated ones, with notable improvements, especially in scenarios when the speech encoder is frozen.
This indicates that HuBERT, which performed representation learning via masked prediction using a pseudo-discrete label, can provide more useful features for the proposed model as a speech encoder compared to wav2vec 2.0, which performed contrastive learning along with auxiliary diversity loss to encourage the use of discrete units.
Moreover, models integrated with HuBERT have shown performance comparable to the ones integrated with Whisper encoder, which has already been trained on a large amount of multi-lingual speech data for ASR tasks, even though pre-trained HuBERT is not fine-tuned in the ASR task.
These results confirmed that integrating HuBERT as the pre-trained speech encoder suits the proposed model framework.

\subsection{Additional results of domain adaptation}
\label{sec:appendix_adapt}

We conducted additional experiments to evaluate the effectiveness of domain adaptation of each module of the proposed model to better understand domain adaptation.
The setting for the adaptation data is the same as in Section \ref{sec:exp_adapt}.

\begin{table}[t]
  \centering
  \resizebox{0.99\linewidth}{!}{%
    \begin{tabular}[t]{cccccccc}
      \toprule
      \multicolumn{3}{c}{\textbf{Adaptation}} & \multicolumn{5}{c}{\textbf{CER [\%] ($\downarrow$)}} \\
      \cmidrule(lr){1-3} \cmidrule(lr){4-8}
      \textbf{\multirowcell{2}{\\[-10pt]HuBERT}} & \textbf{\multirowcell{2}{\\[-10pt]Bridge\\network}} & \textbf{\multirowcell{2}{\\[-10pt]GPT}} & \hspace{-3mm}\textbf{\multirowcell{2}{\\[-10pt]JSUT\\basic5000}}\hspace{-3mm} & \textbf{\multirowcell{2}{\\[-10pt]CV8.0\\test}} & \multicolumn{3}{c}{\textbf{CSJ}} \\
      \cmidrule(lr){6-8}
      & & & & & \textbf{Eval1} & \textbf{Eval2} & \textbf{Eval3} \\
      \midrule\midrule
      \multicolumn{7}{l}{\textbf{No adaptation}} \\
      \quad      &            &            & 8.6 & 9.2 & 30.9 & 26.6 & 22.9 \\
      \midrule
      \multicolumn{7}{l}{\textbf{Adaptation to CV8.0 train (9 hours)}} \\
      \checkmark &            &            & 7.8 & \greycell 9.8 & 30.5 & 26.5 & 23.7 \\
                 & \checkmark &            & 7.7 & \greycell 8.9 & 31.1 & 27.6 & 25.3 \\
                 &            & \checkmark & 7.6 & \greycell \textbf{8.6} & 28.1 & 24.9 & 20.7 \\
      \midrule
      \multicolumn{7}{l}{\textbf{Adaptation to CSJ core (34 hours)}} \\
      \checkmark &            &            & 12.7 & 13.0 & \greycell 17.4 & \greycell 16.9 & \greycell 13.5 \\
                 & \checkmark &            & 9.9 & 13.9 & \greycell 8.0 & \greycell 6.8 & \greycell 4.7 \\
                 &            & \checkmark & 8.7 & 11.2 & \greycell 7.9 & \greycell 7.0 & \greycell 4.9 \\
      \midrule
      \multicolumn{7}{l}{\textbf{Adaptation to CSJ all (520 hours)}} \\
      \checkmark &            &            & 9.2 & 14.6 & \greycell 13.8 & \greycell 13.5 & \greycell 10.3 \\
                 & \checkmark &            & 10.4 & 15.9 & \greycell \textbf{6.4} & \greycell 5.3 & \greycell \textbf{4.0} \\
                 &            & \checkmark & 8.6 & 14.9 & \greycell 6.6 & \greycell 5.3 & \greycell 4.1 \\
      \midrule
      \multicolumn{7}{l}{\textbf{Adaptation to CV8.0 train and CSJ core (43 hours)}} \\
      \checkmark &            &            & 12.1 & \greycell 9.7 & \greycell 17.4 & \greycell 16.5 & \greycell 13.3 \\
                 & \checkmark &            & 8.8 & \greycell 9.4 & \greycell 8.4 & \greycell 7.0 & \greycell 7.0 \\
                 &            & \checkmark & \textbf{7.2} & \greycell 8.9 & \greycell 7.9 & \greycell 6.9 & \greycell 4.7 \\
      \midrule
      \multicolumn{7}{l}{\textbf{Adaptation to CV8.0 train and CSJ all (529 hours)}} \\
      \checkmark &            &            & 9.1 & \greycell 9.8 & \greycell 14.0 & \greycell 13.1 & \greycell 10.4 \\
                 & \checkmark &            & 9.2 & \greycell 9.7 & \greycell 6.6 & \greycell \textbf{5.2} & \greycell 4.1 \\
                 &            & \checkmark & 7.8 & \greycell 9.0 & \greycell 6.6 & \greycell 5.3 & \greycell \textbf{4.0} \\
      \bottomrule
    \end{tabular}
  }
  \caption{
    ASR results with different adaptation settings.
    The gray color means the same domain of test sets as that of adaptation data.
  }
  \label{table:appendix_adapt}
\end{table}

The results are listed in Table~\ref{table:appendix_adapt}.
From the table, domain adaptation to HuBERT alone had a limited effect, unlike the bridge network and GPT.
HuBERT may have been robust to differences in the domain of input speech (e.g., the age group of the target speaker, microphone, and noise conditions) and enables the extraction of domain-independent speech features without being affected by such domain differences.
On the other hand, adapting bridge networks or GPT led to a significant improvement in CER, especially for the adaptation to CSJ data.

While it is intuitively understandable that GPT adaptation plays a crucial role in improving performance for different text domains, the interesting result is that adaptation effective for bridge networks alone is significant.
Bridge networks reduce the temporal resolution of speech representations to be close to the text embedding and embed their representations as speech prompts.
Adaptation to bridge networks helped generate speech prompts from speech representations that would facilitate the generation of text in different domains for subsequent GPT.

\subsection{Mora-level ASR experiments}
\label{sec:appendix_mora}

In order to further evaluate the adaptability of the proposed model to different ASR scenarios, we conducted experiments to adapt the proposed model to Japanese mora-level ASR.
In Japanese, a mora is a rhythmic unit that plays a key role in the language's rhythm and is defined as a unit represented by a single sound of kana and long sound symbol.
Note that the contracted sounds do not form a mora by themselves and are attached to other kana.
All the other of the kana form a mora on their own.
For example, the word ``\jpchar{研究}'' (research in Japanese) is considered to have four morae ``\jpchar{ケ}/\jpchar{ン}/\jpchar{キュ}/\jpchar{ー}'' (ke / N / kyu / --).
In this experiment, the text in the pronunciative form notation contained in the CSJ annotation data was used as the target mora sequence.
We conducted experiments with two different data sizes: core data only (34 hours) and all data (520 hours), as the same in the Section~\ref{sec:exp_adapt}.
The mora set conforms to the CSJ corpus and consists of 145 types.

One approach to extending a pre-trained speech recognition model for mora recognition involves replacing the tokenizer with a vocabulary set of morae. However, this approach may face issues with mismatching the text embedding space, which could prevent leveraging the capabilities of the accumulated language knowledge in the proposed model.
Therefore, as an alternative approach, this experiment treated the mora sequences as plain text, using the same tokenizer before and after adaptation.
In the case of Japanese, the mora sequence can be noted in both katakana and hiragana; for the aforementioned ``\jpchar{研究},'' it is represented as ``\jpchar{ケンキュー}'' in katakana and ''\jpchar{けんきゅー}'' in hiragana.
Hence, both cases were examined in the same configuration except for the mora notation.

\begin{table}[t]
  \centering
  \resizebox{0.88\linewidth}{!}{%
    \begin{tabular}[t]{cccccc}
      \toprule
      \multicolumn{3}{c}{\textbf{Adaptation}} & \multicolumn{3}{c}{\textbf{CER [\%] ($\downarrow$)}} \\
      \cmidrule(lr){1-3} \cmidrule(lr){4-6}
      \textbf{\multirowcell{2}{\\[-7pt]HuBERT}} & \textbf{\multirowcell{2}{\\[-9pt]Bridge\\network}} & \textbf{\multirowcell{2}{\\[-7pt]GPT}} & \multicolumn{3}{c}{\textbf{CSJ}} \\
      \cmidrule(lr){4-6}
      & & & \textbf{Eval1} & \textbf{Eval2} & \textbf{Eval3} \\
      \midrule\midrule
      \multicolumn{6}{l}{\textbf{``Katakana'' mora adaptation to CSJ core (34 hours)}} \\
                 &            & \checkmark & 5.9 & 4.4 & 4.2 \\
                 & \checkmark & \checkmark & 5.1 & 3.6 & 3.4 \\
      \checkmark & \checkmark & \checkmark & 4.8 & 3.7 & 3.2 \\
      \midrule
      \multicolumn{6}{l}{\textbf{``Hiragana'' mora adaptation to CSJ core (34 hours)}} \\
                 &            & \checkmark & 5.7 & 4.2 & 3.9 \\
                 & \checkmark & \checkmark & 4.8 & 3.6 & 3.3 \\
      \checkmark & \checkmark & \checkmark & 4.5 & 3.3 & 3.2 \\
      \midrule
      \multicolumn{6}{l}{\textbf{``Katakana'' mora adaptation to CSJ all (520 hours)}} \\
                 &            & \checkmark & 4.9 & 3.9 & 3.4 \\
                 & \checkmark & \checkmark & 4.4 & 2.8 & 2.6 \\
      \checkmark & \checkmark & \checkmark & 3.9 & 2.7 & 2.6 \\
      \midrule
      \multicolumn{6}{l}{\textbf{``Hiragana'' mora adaptation to CSJ all (520 hours)}} \\
                 &            & \checkmark & 4.6 & 3.3 & 3.0 \\
                 & \checkmark & \checkmark & 4.2 & 2.8 & 2.6 \\
      \checkmark & \checkmark & \checkmark & \textbf{3.7} & \textbf{2.6} & \textbf{2.4} \\
      \bottomrule
    \end{tabular}
  }
  \caption{
    ASR results with different adaptation settings for Japanese Mora-level ASR.
    We used two mora notations, \emph{katakana} and \emph{hiragana}, as the target transcriptions.
  }
  \label{table:appendix_mora}
\end{table}

The results are listed in Table~\ref{table:appendix_mora}.
This table shows that the proposed model adapted to the mora-level transcription achieved a low CER of less than 6.0\%.
In particular, the model adapted only to GPT revealed a performance decrease of only approximately 1\% compared to the model adapted to all modules, suggesting that the proposed model is highly adaptable to various scenarios by simply switching the adaptation module.
It is also interesting to note that ASR systems targeting the mora sequence in hiragana notation had a lower CER than those targeting the mora sequence in katakana notation.
This may be because katakana is more commonly used for foreign words and technical terms, while hiragana is more prevalent in ordinary Japanese texts, including particles and auxiliary verbs.
Therefore, GPT integrated with the proposed model may be more adept at generating text sequences in hiragana.

\end{document}